\documentclass{aa}
\usepackage{psfig}

\def\la{\;
\raise0.3ex\hbox{$<$\kern-0.75em\raise-1.1ex\hbox{$\sim$}}\; }
\def\ga{\;
\raise0.3ex\hbox{$>$\kern-0.75em\raise-1.1ex\hbox{$\sim$}}\; }

\newcommand{\zabs}{$z_{\rm abs}\,$}

\newcommand{\kms}{km~s$^{-1}\,$}
\newcommand{\cm}{cm$^{-2}\,$}

\begin{document}

\title{Spectral energy distribution of the metagalactic ionizing 
radiation field from QSO absorption spectra\thanks{Based
on observations obtained at the VLT Kueyen telescope (ESO, Paranal, Chile),
the ESO programme 65.O-0474(A) 
}
}

\author{
I. I. Agafonova\inst{1}
\and
M. Centuri\'on\inst{2}
\and
S. A. Levshakov\inst{1}
\and
P. Molaro\inst{2}
}

\offprints{S. A. Levshakov}

\institute{Department of Theoretical Astrophysics,
Ioffe Physico-Technical Institute, 194021 St.Petersburg, Russia
\and
Osservatorio Astronomico di Trieste, Via G. B. Tiepolo 11,
34131 Trieste, Italy
}

\date{Received 00  / Accepted 00 }

\abstract{A computational procedure is presented to estimate the 
spectral shape of the ionizing background 
between 1 and 10~Ryd by analyzing optically thin absorption systems
in the spectra of high redshift quasars. 
The procedure is based on the response surface 
methodology from the theory of experimental design.
The shape of the recovered UV background at $z \sim 3$ shows 
a significant intensity decrease between 3 and 4 Ryd
compared to the metagalactic spectrum of Haardt \& Madau (1996).
This decrease is interpreted as produced 
by \ion{He}{ii} Gunn-Peterson effect.
There are no features indicating a contribution from galaxies
to the UV background which is
therefore dominated by QSOs at $z \sim 3$.
}
\authorrunning{I. I. Agafonova et al.}
\titlerunning{SED of the metagalactic ionizing 
radiation field} 
\maketitle

\section{Introduction}

The study of the spectral energy distribution (SED) in the 
radiation background resulting from flux emitted by
all celestial sources
is an important part of modern observational cosmology.
The current paradigm assumes that the metagalactic ionizing background
is formed by radiation of QSOs and galaxies (stars) reprocessed by the 
intergalactic medium (IGM). 
The SED evolves with cosmic time due to
different contribution of galactic and QSO radiation, and 
to varying IGM opacity caused by the hydrogen and helium reionization. 
 
Since the pioneering works of Chaffee et al. (1986) and
Bergeron \& Stasinska (1986), the approach to recover the shape of
the ionizing radiation field 
from the measurements of the intervening metal absorbers in QSO spectra
has been employed in numerous studies. 
In particular, the energy range 1 Ryd $< E <$ 10 Ryd 
can be probed through the relative
intensities of metal lines 
such as \ion{Si}{ii}-\ion{Si}{iv}, \ion{C}{ii}-\ion{C}{iv},
\ion{N}{iii}, \ion{N}{v}, \ion{O}{vi}.
A common procedure consists of selecting a standard SED 
and checking whether it is consistent with measured column densities. 
This procedure can be appropriate for obtaining some 
general information about the SED, but its effectiveness 
and accuracy is rather low because of lack of search strategy. 

In the present paper we describe a computational
technique exploiting the response surface methodology 
from the theory of experimental design 
which enables a {\it directed} search for the shape of the
ionizing background. 
The proposed method reliably recovers
the main features in the spectral shape
of the metagalactic flux by the analysis of
optically thin QSO absorption systems.  
Four absorption systems were selected to
illustrate how this approach can be implemented in practice.

Three of these systems have redshifts $z \sim 3$. 
This redshift is of particular interest for studying the 
traces of still not completely
ionized \ion{He}{ii} (Reimers et al. 1997). In general, the presence of
\ion{He}{ii} in the intergalactic medium
affects the ionizing  spectrum in the range 
$E > 1$ Ryd due to 
\ion{He}{ii} Ly$\alpha$,  two-photon reemission 
and to \ion{He}{ii} continuum absorption (Haardt \& Madau 1996;
Fardal et al. 1998).
The \ion{He}{ii} Ly$\alpha$ absorption 
troughs detected in spectra of five  quasars (Vogel \& Reimers 1995;
Zheng et al. 1998; Anderson et al. 1999; Heap et al. 2000; Smette et al. 2002;
Jakobsen et al. 2003 ; Zheng et al. 2004a)
may indicate \ion{He}{ii} Gunn-Peterson effect
(\ion{He}{ii} Ly$\alpha$ absorption in a diffuse IGM)
at $z \sim 3$. 
Furthermore, recent FUSE observations of the \ion{He}{ii} Ly$\alpha$ 
forest towards HE~2347--4342 combined with observations of the
\ion{H}{i} Ly$\alpha$ forest at the VLT revealed
also large scale variations in $\eta = \ion{He}{ii}/\ion{H}{i}$ 
which are still not well understood (Shull et al. 2004; Zheng et al. 2004b). 
One possible explanation is the presence of spatial fluctuations in the 
metagalactic ionizing field at $E \ga 4$ Ryd. 
To confirm this suggestion, additional observations
at different redshifts and in different sightlines are needed. 
However,  direct measurements 
of \ion{He}{ii} Ly$\alpha$ absorption at
303.78 \AA\, are very problematic due to almost complete 
light blotting of distant QSOs by the intervening Lyman limit
or damped Ly$\alpha$ systems (LLSs and DLAs, respectively). 
In this context, the proposed approach to
estimate the shape of the ionizing background from metal absorption
systems is of great importance.   
 
The structure of the paper is as follows. 
The procedure to recover the shape of the underlying UV 
continuum and an example illustrating how to use it  
are described in Sect.~2. This section also contains a brief description of a
computational method used to invert the observed line profiles -- 
the Monte Carlo Inversion (MCI).  
The detailed analysis of absorption systems used 
for the SED estimations is given in Sect.~3. 
The results are discussed and summarized in Sect.~4. 
An example of an experimental design is given in the Appendix.

\section{Computational methods}

\subsection{Shape of the ionizing radiation}

Here follows a description of how 
the SED of the ionizing radiation can be estimated from
metal line profiles observed in optically thin absorption
systems. 
The method is based on the 
response surface methodology used in experimental design
(see, e.g., Box, Hunter \& Hunter 1978, Chapter 15).

A basic UV spectrum is taken as an initial guess.
The estimated column densities of different ions
and a photoionization code (e.g., CLOUDY, Ferland 1997)  
are used to derive the ionization parameter,
metallicity and element abundance ratios.
If the observed column densities 
are well reproduced with this trial spectrum, 
it implies that the initial guess was appropriate.
If, however, the observed column densities
cannot be reproduced or/and some other inconsistencies arise, then 
the assumed shape of the UV background is to be adjusted.  

To allow for quantitative estimations, the shape of the UV continuum  
must be parameterized.
In general, the shape of the ionizing spectrum can be specified by 
a piecewise continuous function,
e.g. by a set of power laws and/or exponents.
In particular,
in the range 1 Ryd $< E < 10$ Ryd relevant for the ions 
frequently observed in
QSO spectra, the following variables (called `factors' in the 
experimental design) can be used 
to describe the main features of the SED (Fig.~1): 

A first region between 1~Ryd and the \ion{He}{i} break is
characterized by a slope $f_1$ (power law exponent)  
and by the position $f_2$ of the \ion{He}{i} break.
A second region can be defined between the \ion{He}{i} 
and \ion{He}{ii} breaks (point $A$) with $f_3$ and
$f_4$ the slope and the position of the \ion{He}{ii} break, respectively.
Since the energy depression can be relevant, we introduce here also
$f_5$ which defines the depth of the break
[$f_5 = \log (J_{\rm B}/J_{\rm A})$], and 
$f_6$~-- the slope between points $A$ and $B$.
$f_7$ characterizes the energy spectrum in the region $B-C$.
To account for a possible bump around 3 Ryd due to
recombinations within the clumpy intergalactic gas (\ion{He}{ii} Ly$\alpha$ 
emission, \ion{He}{ii} two-photon continuum emission and \ion{He}{ii}
Balmer continuum emission), we introduce factors 
$f_8$ and $f_9$ which describe the amplitude of the bump  
and its width, respectively.  

\begin{figure}[t]
\vspace{0.0cm}
\hspace{0.0cm}\psfig{figure=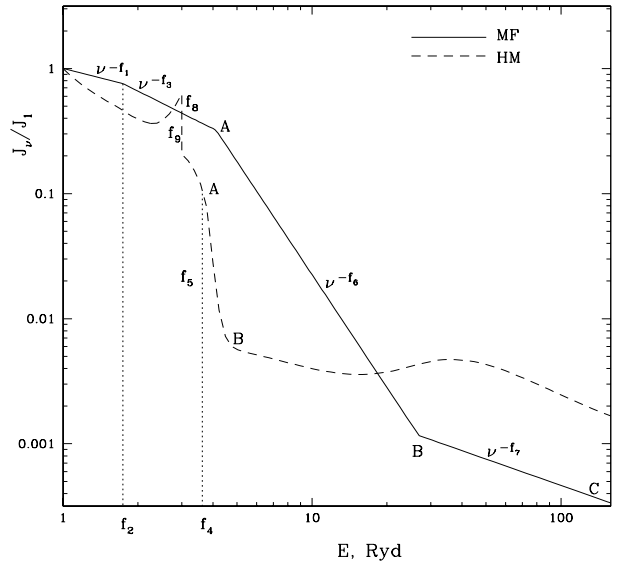,height=8.0cm,width=8.0cm}
\vspace{-0.4cm}
\caption[]{Schematic picture of the $z=3$ metagalactic (dashed line) and
AGN-type (smooth line) ionizing backgrounds from  
Haardt \& Madau (1996) and Mathews \& Ferland (1987), respectively.
The spectrum is normalized so that $J_\nu(h\nu =$ 1 Ryd) = 1.
The emission bump at 3 Ryd in the metagalactic spectrum
is caused by reemission of \ion{He}{ii} Ly$\alpha$, \ion{He}{ii}
two-photon continuum emission and \ion{He}{ii} Balmer continuum emission
from intergalactic clouds.
The definition of factors $\{f_i\}$ is given in Sect.~2.1
}
\label{fig1}
\end{figure}

The energy of the far UV cut off (point $C$) 
when taken above 100 Ryd does not 
affect the fractional ionizations of ions we are interested in. 
In all computations described in subsequent sections
this energy and the slope after it are kept fixed at 128 Ryd and --1.5, 
respectively.
Note that the positions of the \ion{He}{i} and \ion{He}{ii} breaks
can be shifted due to large scale motions and/or superposition of different
spectra (e.g. stellar+metagalactic), and hence, in general the values of 
$f_2$ and $f_4$ may not be equal to 
1.8 Ryd and 4 Ryd, respectively.
Additional factors can be defined to describe some particular 
features of the SED as, for example, those caused by
\ion{He}{ii} Ly$\alpha$ absorption in a diffuse IGM 
(see Sect.~3 below). 
 
In this notation, an ionizing spectrum is determined as a point
$\{f_1, \ldots, f_k\}$ in the $k$-dimensional factor space. 
The number and selection of factors depend on a particular
spectral shape.  For instance, 
the metagalactic spectrum at $z = 3$ calculated by 
Haardt \& Madau (1996, hereafter HM) has coordinates 
$\{f_1, f_4, f_5, f_6, f_7, f_8, f_9\} =$
$\{$ $-1.4$, 0.55, $-1.3$, $-13.0, -0.45$, 0.46, 0.11$\}$
with $f_1$ standing for the entire slope between 1 Ryd and $f_4$,
whereas the point 
$\{f_1, f_2, f_3, f_4, f_5, f_6, f_7\} =$
$\{$ $-0.5$, 0.25, $-1.0$, 0.61, $-2.5$, $-3.0, -0.7$ $\}$ corresponds to
the AGN-type spectrum of Mathews \& Ferland (1986, hereafter MF) 
(all energies are given in $\log_{10}$). 

\begin{figure}
\vspace{-0.4cm}
\hspace{-0.5cm}\psfig{figure=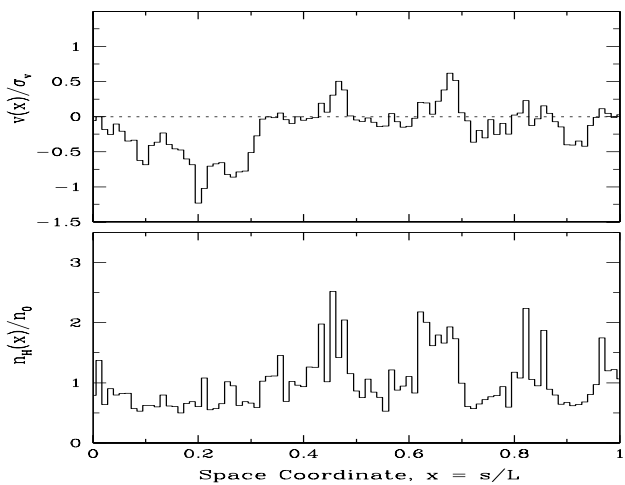,height=8.0cm,width=12.0cm}
\vspace{-1.4cm}
\caption[]{
Velocity (upper panel) and gas density (lower panel) distributions
along the line of sight used to produce the mock absorption spectra
which are shown in Fig.~4
}
\label{fig2}
\end{figure}

The next step of the shape adjustment procedure is to find 
a direction in the factor space which leads  
to the UV background with more appropriate characteristics.
The starting (`null') point is represented by the initial spectrum.
A set of new trial spectra is produced
by varying the factor values around the `null' point
in accordance with a special scheme called `experimental (factorial) 
design'. The experimental design is represented by
a matrix with $n$ ($n > k$) rows containing particular values of factors 
(an example of the experimental design is given in Appendix).

To evaluate the fitness of each trial ionizing spectrum 
a numerical measure $\tilde{{\cal R}}$ (usually called `response')
is to be defined in such a way that bigger values of $\tilde{{\cal R}}$ 
correspond to increasing fitness. In general,
the choice of $\tilde{{\cal R}}$ occurs rather heuristically and 
accounts for the information obtained with the initial UV spectrum
and for any a priori information (like, e.g., allowable element abundance
ratios). For illustration consider the following two 
examples. 

Assume that a set of silicon 
(\ion{Si}{ii}, \ion{Si}{iii}, \ion{Si}{iv})
and carbon (\ion{C}{ii}, \ion{C}{iii}, \ion{C}{iv}) lines is observed in
an absorption system, and that the initial UV spectrum 
overestimates
the column densities of 
\ion{Si}{ii} and \ion{C}{iv} and underestimates those of
\ion{Si}{iv} and \ion{C}{ii}. 
Then the response can be written as
$$
\tilde{{\cal R}} =  \frac{\ion{C}{ii}}{\ion{C}{iv}} \times 
\frac{\ion{Si}{iv}}{\ion{Si}{ii}},
$$
i.e., the search should go towards increasing the product of
these ratios. 

Another example concerns an absorption system with the observed lines
of \ion{Si}{iii}, \ion{Si}{iv}, and \ion{C}{iv}.
The ratio \ion{Si}{iii}/\ion{Si}{iv} determines the ionization parameter $U$.
Using the corresponding ion fractions $\Upsilon_i$    
the ratio
$$
\frac{\rm Si}{\rm C} = \frac{\ion{Si}{iv}}{\ion{C}{iv}}\,
\frac{\Upsilon_{\rm C\,IV}}{\Upsilon_{\rm Si\,IV}}   
$$
can be calculated. Both observations and theoretical considerations
give for this ratio a safe upper limit
Si/C $< 3$(Si/C)$\odot$.
If, for instance, the initial UV background delivers Si/C = 10(Si/C)$\odot$,
then the search for a new SED can be governed by the ratio
$$
\tilde{{\cal R}} = \frac{\Upsilon_{\rm Si\,IV}}{\Upsilon_{\rm C\,IV}}\; ,
$$ 
calculated for the ionization parameter $U$ given by the observed
column density ratio
\ion{Si}{iii}/\ion{Si}{iv} (note that the value of $U$ depends on the
shape of the trial spectrum).
More examples are presented  in the subsequent sections. 

\begin{figure}[t]
\vspace{0.0cm}
\hspace{0.0cm}\psfig{figure=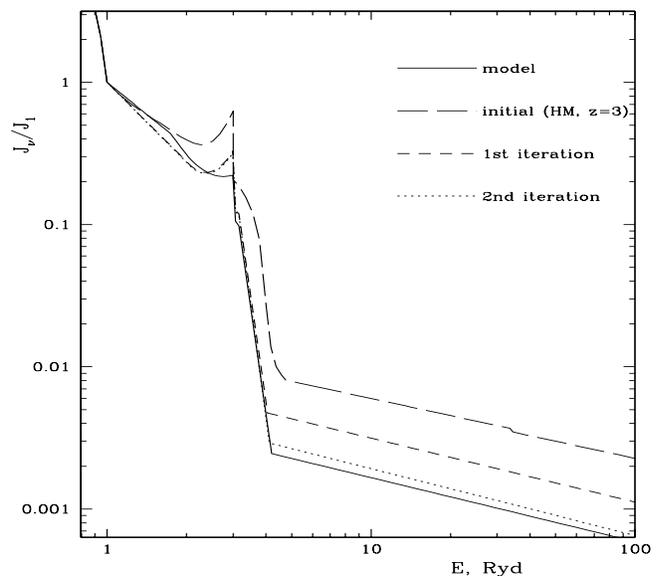,height=8.0cm,width=8.0cm}
\vspace{-0.4cm}
\caption[]{
A model ionizing background (smooth line) used to calculate
the mock spectra shown in Fig.~4. As an initial approximation
for the restoring procedure an HM-type spectrum is taken (long-dashed line). 
The results of the first and second iterations
are illustrated by the short-dashed and dotted lines, respectively 
}
\label{fig3}
\end{figure}

The calculated responses  $\{\tilde{{\cal R}}_i\}^n_{i=1}$ 
can be considered as points
belonging to some surface in $k$-dimensional factor space (referred
to as `response surface'). 
To determine the direction of steepest ascent the response surface should be
described by some analytical function.  
A standard linear model is a reasonable first approximation:
\begin{equation}
{\cal R} = \sum^k_{i=1}\,\alpha_i \hat{f}_i + \beta\, , 
\label{eq:E1b}
\end{equation}   
where $\{ \alpha_i \}^k_{i=1}$ are factor effects,  $\hat{f}_i$ is the 
scaled and centered value of
the $i$th factor, $\hat{f}_i = (f_i - f_{0,i})/s_i$,
$s_i$ is some suitable scale (range of variation) of the $i$th factor,
and $\beta = \alpha_0 + \sum^k_{i,j=1}\,\alpha_{i,j}
\hat{f}_i\hat{f}_j$
stands for the joint effect of the free term $\alpha_0$ and of the
nonlinear term that contains also the interaction of factors.

\begin{figure*}[t]
\vspace{0.5cm}
\hspace{-0.5cm}\psfig{figure=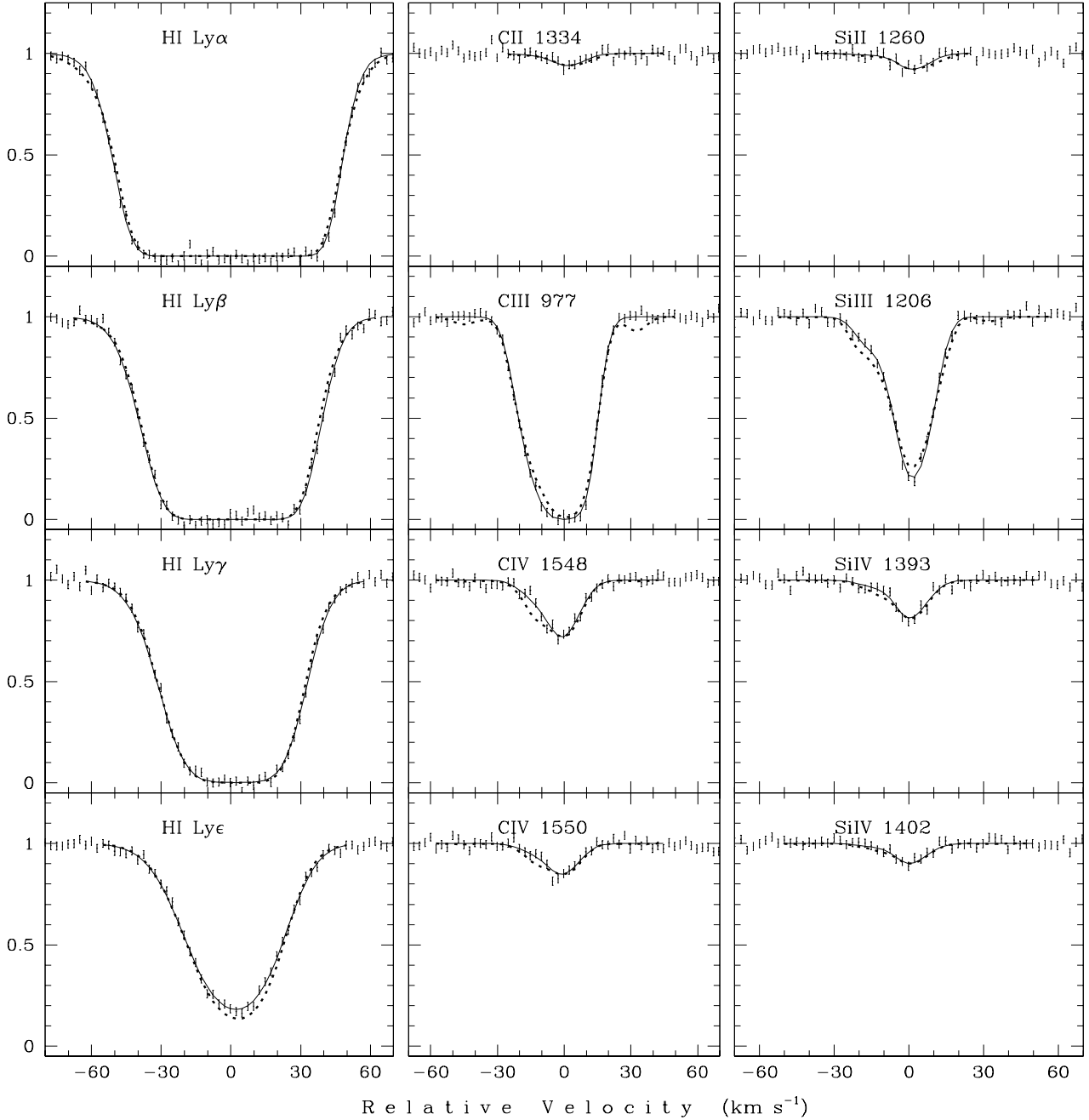,height=18.0cm,width=19.0cm}
\vspace{-0.3cm}
\caption[]{Mock spectra of hydrogen and metal absorption lines 
(dots with 1 $\sigma$ error bars). 
The MCI solution corresponding to the initial ionizing spectrum 
is shown by the dotted curves.
The smooth curves are the synthetic profiles obtained with recovered
2nd iteration SED (dotted line in Fig.~3)
}
\label{fig4}
\end{figure*}

The coefficients $\alpha_i$ and $\beta$ and their dispersions are calculated 
from $n$ values of ${\cal R}$  according to special formulas 
(depending on the type of experimental design used). 
The validity of the planar data model (\ref{eq:E1b})
can be checked by comparison of the estimation, $\beta'$, 
with the value of ${\cal R}_0$ at the `null point', i.e.,
when $f_i = f_{0,i}$. Statistical
significance of the difference (${\cal R}_0 - \beta'$) points to
non-negligible non-linear effects.
In this case the higher-order data model  
is to be used with corresponding higher-order experimental design.

\begin{table*}[t]
\centering
\caption{Physical parameters for model absorption system shown in Fig.~4}
\label{tbl-A1}
\begin{tabular}{lcccc}
\hline
\noalign{\smallskip}
Parameter & Model & Estimated with & Estimated with & Estimated with \\
          & values& initial SED & 1st iteration SED & 2nd iteration SED\\
(1) & (2) & (3) & (4) & (5) \\
\noalign{\smallskip}
\hline
\noalign{\smallskip}
$U$ & 1.55E-2 & 5.7E-2 & 1.3E-2 & 1.4E-2\\
$N_{\rm H}$, \cm & 1.2E19 & 2.5E18 & 8.4E18 & 1.0E19\\
$\sigma_{\rm v}$, \kms & 20.0 & 17.1 & 22.2 & 18.0\\
$\sigma_{\rm y}$ & 0.40 & 0.56 & 0.50 & 0.48\\
$Z_{\rm C}$ & 8.0E-6 & 2.7E-5 & 1.0E-5 & 9.0E-6\\
$Z_{\rm Si}$ & 1.0E-6 & 3.8E-6 & 1.6E-6 & 1.2E-6\\
$N$(\ion{H}{i}), \cm & 3.4E15 & 3.7E15 & 3.4E15 & 3.4E15\\
$N$(\ion{C}{ii}), \cm & 2.5E12 & 3.0E12 & 2.9E12 & 2.6E12\\
$N$(\ion{C}{iii}), \cm & 8.1E13 & 5.5E13 & 6.9E13 & 7.7E13\\
$N$(\ion{C}{iv}), \cm & 9.0E12 & 9.9E12 & 9.2E12 & 8.6E12\\
$N$(\ion{Si}{ii}), \cm & 3.5E11 & 4.2E11 & 3.9E11 & 3.5E11\\
$N$(\ion{Si}{iii}), \cm & 5.5E12 & 5.2E12 & 5.4E12 & 5.3E12\\
$N$(\ion{Si}{iv}), \cm & 1.9E12 & 2.2E12 & 1.8E12 & 1.9E12\\
\hline
\noalign{\smallskip}
\multicolumn{5}{l}{Note: $Z_{\rm X} = N_{\rm X}/N_{\rm H}$}\\
\end{tabular}
\end{table*}

Given the factor effects $\{ \alpha_i \}$,
the  maximization of ${\cal R}$ is obtained  by moving the factor values from
the `null' point 
in the direction normal to the response surface. 
This movement is performed stepwise until either 
the desired UV background is found
(i.e. the one which reproduces self-consistently 
all observed column densities) 
or the validity limit of the employed data
model is reached. In the latter case the adjustment procedure
should be repeated, this time with a newly obtained UV spectrum as a
`null' point. 

The initial set of factors may turn out to be redundant, i.e. some of the
factor effects $\alpha_i$ may be statistically insignificant. 
This means that the corresponding features in the spectral shape
do not affect the absorption line  profiles included in the analysis.
Such insignificant factors are removed
from the data model (\ref{eq:E1b}) and are fixed at some appropriate level.
The effects of the remaining $m$ factors are then recalculated using
the corresponding $m$-factorial design.

\begin{figure*}[t]
\vspace{0.4cm}
\hspace{-0.5cm}\psfig{figure=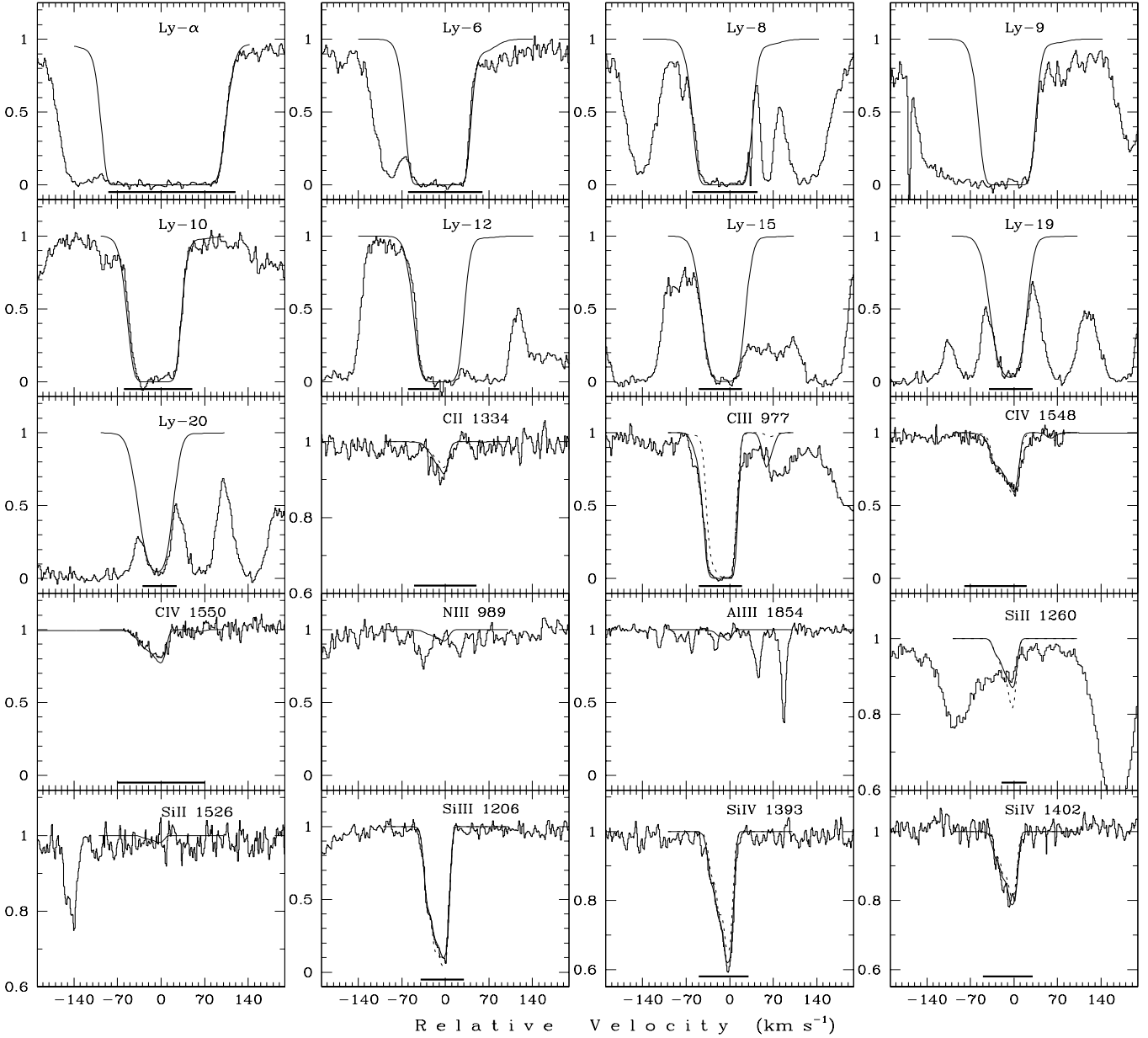,height=17.0cm,width=19.0cm}
\vspace{-0.5cm}
\caption[]{
Hydrogen and metal absorption lines associated with the \zabs = 2.9171 system
towards \object{HE 0940--1050} (histograms).
Synthetic profiles corresponding to the MCI solution with the 
$z=3$ HM ionizing spectrum (smooth line in Fig.~6)
are shown by the dotted curves, whereas those calculated with the final adjusted
SED (short-dashed line in Fig.~6)
are the smooth curves. The corresponding physical parameters are listed 
in Table~2, Col.~2.
Bold horizontal lines mark pixels included in the optimization procedure.
The zero radial velocity is fixed at $z = 2.9171$ 
}
\label{fig5}
\end{figure*}

The uncertainty of the recovered spectral shape requires some comments.
The spectrum of the ionizing radiation, $J_\nu$, defines the
rates (s$^{-1}$) of the photoionization processes, $\Gamma_{ij}$,
through the integral
\begin{equation}
\Gamma_{ij} = \int^\infty_{\nu_{ij}}\,
\frac{4\pi J_\nu}{h\nu}\,\sigma_{ij}(\nu)\,d\nu\; ,
\label{eq:E1a}
\end{equation}   
where $\sigma_{ij}(\nu)$ is the photoionization cross section at
frequency $\nu$ for species $i$ in ionization state $j$, and
$\nu_{ij}$ is the threshold frequency for photoionization.

Thus, the SED is obtained by solving the integral equation 
which is a classic ill-posed problem. Its solution depends crucially
from the number of constraints included in the analysis. 
When applied to our problem, this means
that the SED is best recovered if the corresponding absorption system
reveals many lines of different metals in different ionization stages. If
only a few metal lines are observed, they can be used to recover
the shape not in the entire range 1 Ryd $< E < 10$ Ryd, but in some narrower
regions. For example, the lines of \ion{C}{iii}, \ion{C}{iv} and 
\ion{O}{vi} allow to estimate the SED in the vicinity of the \ion{He}{ii} 
break (3 Ryd $< E < 4.5$ Ryd).

\subsection{The Monte Carlo Inversion procedure}

Absorption systems are analyzed by means of 
the Monte Carlo Inversion (MCI) procedure 
described in detail in Levshakov, Agafonova \& Kegel 
(2000, hereafter LAK), and with modifications in 
Levshakov et al. (2002, 2003a,b).
Here we briefly outline the basics
needed to understand the results presented below in Sect.~3. 

The MCI is based on the assumption 
that all lines observed in the absorption system are formed in
a continuous medium 
where the gas density, $n_{\rm H}(x)$, and velocity, $v(x)$,
fluctuate from point to point giving rise to complex profiles 
(here $x$ is the space coordinate along the line of sight).

The MCI also assumes 
that within the absorber the metal abundances are constant,
the gas is optically thin for the ionizing UV radiation, and the gas
is in thermal and ionization equilibrium.
The intensity and the spectral shape of the background ionizing
radiation are treated as external parameters.

The radial velocity  $v(x)$ and gas density $n_{\rm H}(x)$
are considered as two continuous random functions which are
represented by their sampled values at equally spaced intervals
$\Delta x$. The computational procedure uses the
adaptive simulated annealing. The fractional ionizations of
all elements included in the analysis are computed at every space coordinate $x$
with the photoionization code CLOUDY.

In the MCI procedure the following physical 
parameters are directly estimated:
the mean ionization parameter $U_0$,
the total hydrogen column density $N_{\rm H}$,
the line-of-sight velocity dispersion $\sigma_{\rm v}$, and
density dispersion $\sigma_{\rm y}$, of the bulk material
[$y \equiv n_{\rm H}(x)/n_0$],
and the chemical abundances $Z_{\rm a}$ of all elements
involved in the analysis.
With these parameters we can further calculate
the column densities $N_{\rm a}$ for different species, and
the mean kinetic temperature $T_{\rm kin}$. 
If the absolute intensity of the UV background is known, 
then the mean gas number density $n_0$, and the line-of-sight thickness $L$ 
of the absorber can be evaluated as well.

In general, the uncertainties of the fitting parameters $U_0$, $N_{\rm H}$, 
$\sigma_v$, $\sigma_y$, and $Z_a$ are about 15\%--20\% 
(for data with S/N $\ga 30$)
and the errors of the estimated column densities are less than 10\%. 
However, in individual absorption systems, the accuracy of recovered values can
be lower due to different reasons such as partial blending of line profiles,
saturation of profiles or absence of lines of subsequent ionic transitions.

The procedure of spectral shape adjustment described in the preceding section 
is used in a modified form.
From computational point of view, it is more convenient within the MCI
to evaluate responses $\tilde{{\cal R}}$ 
using ion fractions rather then column densities
(at a given metallicity, the
column density of an ion is proportional to its fraction). 
The procedure stops when spectral shape ensuring $\chi^2 \la 1$ for all
lines observed in the system is found.

\begin{figure}[t]
\vspace{0.0cm}
\hspace{0.0cm}\psfig{figure=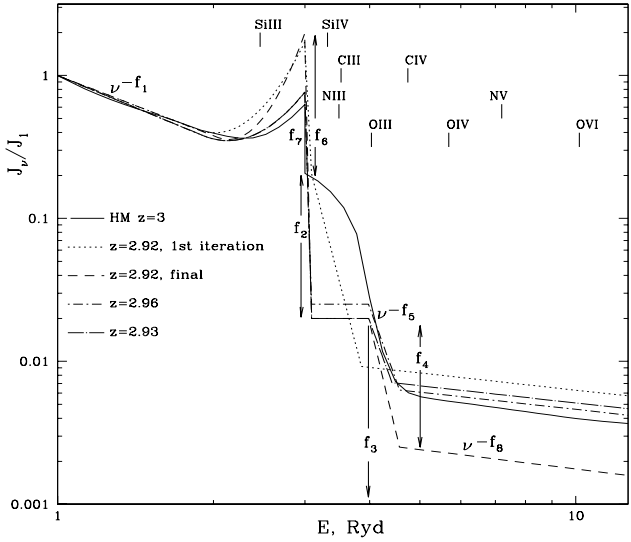,height=8.0cm,width=8.0cm}
\vspace{-0.4cm}
\caption[]{SEDs recovered from the absorption line systems. 
The positions of ionization thresholds of different ions are
indicated by tick marks. The definition of factors $\{f_i\}$ are
given in Sect.~3.1. 
For details, see text }
\label{fig6}
\end{figure}

\subsection{Testing the SED recovering procedure} 

In order to test the procedure of the SED adjustment 
we prepared a mock absorption line spectrum using
the density and velocity distributions
plotted in Fig.~2, the UV ionizing background shown  
in Fig.~3 (solid line), and the model
parameters listed in Table~1, Col.~2. 
The resulting absorption lines after convolution of the intensities  
with a Gaussian-type point-spread
function of FWHM = 7 \kms\, and the addition of white noise with dispersion 0.02
(S/N = 50) are shown in Fig.~4 by points with error bars.
The chosen UV background can
be produced by a mixture of stellar and metagalactic spectra
(see, e.g., Giroux \& Shull 1997).

The procedure requires an input SED and as  
initial guess for the underlying UV background we took the
metagalactic spectrum of HM shown as a
long-dashed line in Fig.~3. 
This spectrum was defined by 7 factors  
$\{$$f_1, f_4, f_5, f_6, f_7, f_8, f_9$$\}$
with the meanings described in Sect.~2.1. 
The only change is for
$f_1$ which now stands for the entire slope between
1 Ryd and the \ion{He}{ii} break at $f_4$.

The best MCI solution gives the physical parameters listed
in Table~1, Col.~3. The corresponding synthetic absorptions
are shown in Fig.~4 by the dotted lines. 
As expected, not all absorptions are well reproduced,
since the tried UV background differs from the one
used to generate the mock lines. 
An underestimation of the
\ion{C}{iii} and \ion{Si}{iii} intensities along with overestimation of
\ion{C}{iv} is well seen in Fig.~4.

To calculate responses, the following expression was chosen:  
$$
\tilde{{\cal R}} = \log\left(\frac{\Upsilon_{\rm C\,III}}
{\Upsilon_{\rm C\,IV}}\, 
\frac{\Upsilon_{\rm Si\,III}}{\Upsilon_{\rm Si\,II}}\right)\, , 
$$
with an additional constraint
$\log (\Upsilon_{\rm Si\,III}/\Upsilon_{\rm Si\,IV}) \ga 0.4$  
[i.e. the fractions of the corresponding ions are calculated for
the value of $U$ determined by the condition
$\log (\Upsilon_{\rm Si\,III}/\Upsilon_{\rm Si\,IV}) \ga 0.4$].

Factors were varied at 2 levels according to a 7-factorial saturated
simplex design which gave 8 new trial UV spectra (see Appendix). 
These spectra were
inserted into CLOUDY and with the obtained fractional ionizations 
the responses for every spectrum were evaluated. 
Then the factor effects were estimated, linearity of data model 
checked, and a
new UV background produced by moving the factor values in the 
direction normal to the response surface.  

The new solution for the background
is shown in Fig.~3 by the short-dashed line, and
the results of the MCI calculations are given in Table~1, Col.~4. 
Fitting of synthetic profiles to most absorption lines becomes much better, 
but \ion{C}{iii}, \ion{C}{iv} and \ion{Si}{iv} lines still have large $\chi^2$ 
values.
Thus, the adjustment procedure was repeated once more,
with the same set of factors, but different response 
$$
\tilde{{\cal R}} = \log\left(\frac{\Upsilon_{\rm C\,IV}}{\Upsilon_{\rm C\,II}}\,
\frac{\Upsilon_{\rm Si\,IV}}{\Upsilon_{\rm Si\,II}} \right)\, ,
$$
with constraints
$\log (\Upsilon_{\rm C\,III}/\Upsilon_{\rm C\,IV}) >$ 0.9 and
$\log (\Upsilon_{\rm Si\,III}/\Upsilon_{\rm Si\,IV}) >$ 0.5.

The resulting UV spectrum is shown in Fig.~3 by dots, and the corresponding
synthetic profiles by the smooth lines in Fig.~4. The recovered parameters are
listed in Col.~5, Table~1. Now all absorption lines are 
well reproduced,
and, thus, the inverse procedure can be considered  completed.

This test demonstrates both the effectiveness of the adjustment
procedure and its limitations. Starting from some standard spectral shape, we
were able to recover quite correctly all significant features of the underlying
ionizing spectrum such as the depth of the 
\ion{He}{ii} break and its shift to the lower energies. 
On the other hand, some fine features of the underlying spectrum
(break at 1.8 Ryd, small bump at 3 Ryd) cannot be reproduced with
the adopted noise level.
Higher S/N data would be more appropriate for this case.

\begin{figure*}[t]
\vspace{0.4cm}
\hspace{-0.5cm}\psfig{figure=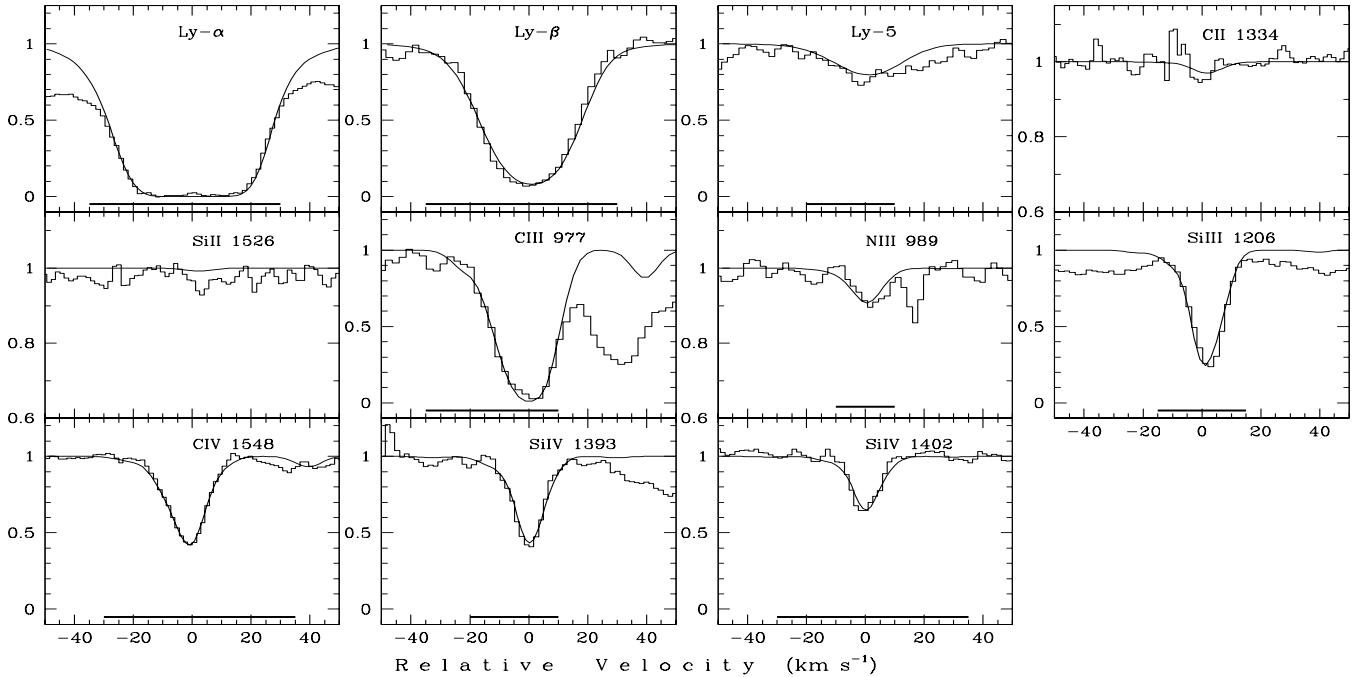,height=15.0cm,width=18.5cm}
\vspace{-6.0cm}
\caption[]{
Hydrogen and metal absorption lines associated with the \zabs = 2.9659 system
towards \object{Q 0347--3819} (histograms).
Synthetic profiles corresponding to the MCI solution for  
the final adjusted SED (dot-short-dashed line in Fig.~6)
are shown by the smooth curves.
The corresponding physical parameters are listed 
in Table~2, Col.~3.
Bold horizontal lines mark pixels included in the optimization procedure.
The zero radial velocity is fixed at $z = 2.9659$ 
}
\label{fig7}
\end{figure*}

\section{Application to observed absorption systems}

Here we analyze some QSO absorption systems using
the procedure described in the preceding section.
All computations below were performed with laboratory wavelengths and 
oscillator strengths taken from Morton (2003). Solar abundances 
were taken from Asplund, Grevesse \& Sauval (2005). 

\subsection{Absorption system at \zabs = 2.9171 towards \object{HE 0940--1050}}

This absorption system shown in Fig.~5
exhibits higher order \ion{H}{i} Lyman series lines 
and transitions of \ion{C}{ii}/\ion{C}{iii}/\ion{C}{iv}, and 
\ion{Si}{ii}/\ion{Si}{iii}/\ion{Si}{iv}.

The system was described in detail in  
Levshakov et al. (2003c) as an extremely low metallicity LLS with 
$N$(\ion{H}{i}) $= 3\times 10^{17}$ \cm\, 
and $Z = 0.001\,Z_\odot$. 
Levshakov et al. also noted that the observed line intensities
were inconsistent with the HM metagalactic ionizing spectrum, and 
that better fitting of the observed profiles corresponded to
the ionizing spectrum having a significantly enhanced
\ion{He}{ii} re-emission bump at 3 Ryd. 
This result was obtained by common `trials and errors' method.
Here we re-analyze the \zabs = 2.9171 system using the procedure of
the directed search.

With the HM ionizing spectrum at $z = 3$ as initial guess 
we obtained the
synthetic profiles shown in Fig.~5 by dotted curves. 
Strong overestimation of \ion{Si}{ii} $\lambda1260$ \AA\,
is clearly seen along with underestimation of 
\ion{C}{ii} $\lambda1334$ \AA\, and \ion{Si}{iv} $\lambda1393$ \AA.
Whether \ion{C}{iii} $\lambda977$ \AA\, line is underestimated or not, 
cannot be decided unambiguously at this stage of investigation
since the apparent intensity may be due 
to contamination by some Ly$\alpha$ interloper. 
The HM spectrum was parameterized using 
the same 7 factors as in the example
described in Sect. 2.3. 
The responses were calculated in the form  
$$
{\cal R} = \log\left( \frac{\Upsilon_{\rm C\,II}}{\Upsilon_{\rm C\,IV}}\, 
\frac{\Upsilon_{\rm Si\,IV}}{\Upsilon_{\rm Si\,II}} \right). 
$$ 
The design employed was the 7-factorial saturated simplex design 
(see Appendix). 
The first-iteration ionizing spectrum 
revealed a sharp \ion{He}{ii} break shifted to 3 Ryd and  the 
\ion{He}{ii} re-emission bump significantly increased
as compared to the initial approximation (Fig.~6, dotted line).
This spectrum provided better fitting to the observed profiles 
(e.g., all carbon
lines including \ion{C}{iii} were now described with $\chi^2 \sim 1$), 
but the
profiles of \ion{Si}{ii} and \ion{Si}{iv} remained, 
correspondingly, over- and underestimated. 

\begin{figure*}[t]
\vspace{0.4cm}
\hspace{-0.5cm}\psfig{figure=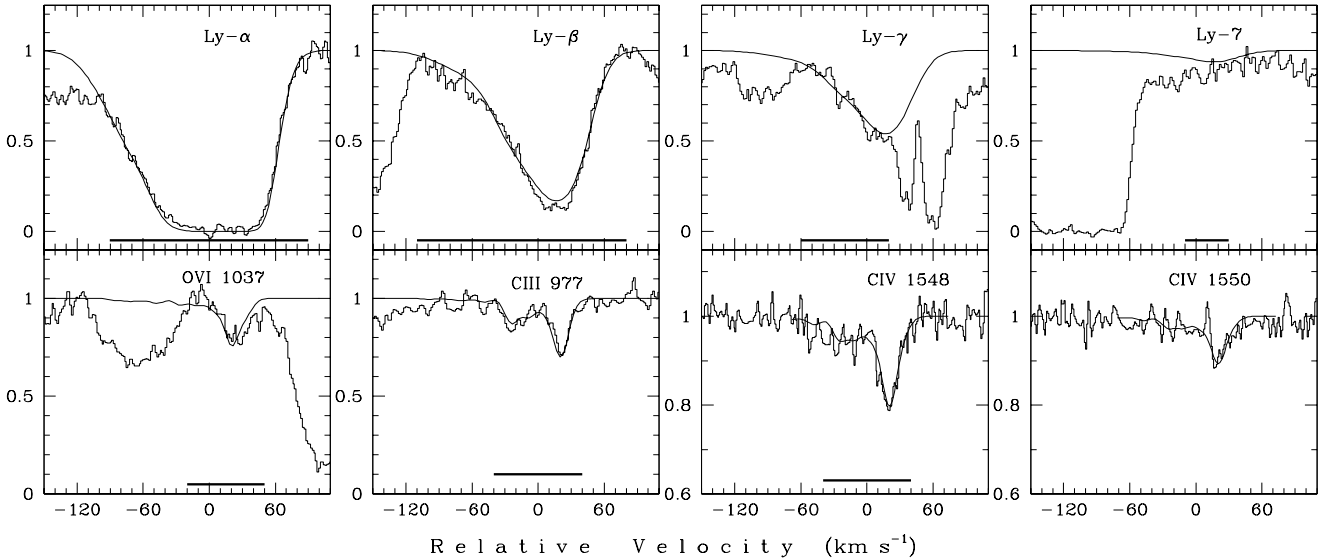,height=13.0cm,width=18.5cm}
\vspace{-5.5cm}
\caption[]{
Same as Fig.~7 but for the \zabs = 2.9375 system towards
\object{HE 0940--1050}.
The MCI synthetic profiles (smooth curves) are calculated with
the final adjusted SED shown by the dot-long dashed line in Fig.~6.
The corresponding physical parameters are listed 
in Table~2, Col.~4.
The zero radial velocity is fixed at $z = 2.9375$ 
}
\label{fig8}
\end{figure*}

Next iteration of the SED adjustment was performed 
with the same set of factors,
design and response augmented with constraints
$\log (\Upsilon_{\rm C\,III}/\Upsilon_{\rm C\,IV}) < 1$ and
$\log(\Upsilon_{\rm Si\,III}/\Upsilon_{\rm Si\,IV}) > 0.1$ to enable
self-consistent description
of the \ion{C}{iii} and \ion{Si}{iii} lines.
However, this step failed to produce any improvement.
Attempts to include additional factors describing 
the \ion{He}{i} break (see Fig.~1) or
to employ other forms of the responses failed as well.
 
The solution was found after performing model 
calculations (using CLOUDY) of radiation
transmission through a  plane-parallel absorbing cloud.
For an HM-type ionizing spectrum, 
$\eta$ = $N$(\ion{He}{ii})/$N$(\ion{H}{i})~$> 50$ 
and the absorber with $N$(\ion{H}{i}) $= 3\times 10^{17}$ \cm,  
marginally optically thick in hydrogen continuum,
is certainly opaque in \ion{He}{ii}. 
The radiation in the
\ion{He}{ii} continuum is effectively absorbed by such cloud 
and a part of this radiation is reemitted
as the \ion{He}{ii} $\lambda304$ \AA\, Ly$\alpha$ 
and two-photon reemission,  and 
\ion{He}{ii} Balmer continuum emission.
As a result, the incident HM spectrum 
being transmitted through the absorber with 
$N$(\ion{H}{i}) $= 3\times 10^{17}$ \cm\,  reveals a sharp 
and deep break at 4 Ryd and a strong emission
line at 3 Ryd of about twice the incident intensity. 
Thus, the increased intensity at $E \la 3$ Ryd in the ionizing spectrum
recovered from the \zabs = 2.92 absorber is, probably, due 
to auto-emission of the cloud. 

However,
in the recovered spectrum a strong intensity depression starts 
just above 3 Ryd, and
this result has been reproduced in all trials with different 
responses and different sets of factors.
At the same time, the model calculations 
show that the spectral range 
3 Ryd $< E < 4$ Ryd remains unaffected by the processes inside the
cloud, i.e., 
if the intensity depression in this range 
is not present in the incident spectrum, 
then it does not appear in the transmitted one. 
This means that the depression at $E \ga 3$ Ryd in the ionizing spectrum 
is produced by the processes outside the cloud. 
We can assume that this is a  manifestation of  
\ion{He}{ii}  Ly$\alpha$ $\lambda304$ \AA\, absorption
arising in both  
smoothly distributed IGM (\ion{He}{ii} Gunn-Peterson effect)
and discrete Ly$\alpha$ forest clouds (line blanketing).   
\ion{He}{ii} Ly$\alpha$ absorption in the diffuse IGM gas results in 
the absorption trough blueward of the resonant wavelength 304~\AA\, (3 Ryd).
This trough together with the subsequent absorption in the
\ion{He}{ii} continuum ($E \ge 4$ Ryd) produce a winding structure in the
spectral shape at $E \ga$ 3 Ryd.
As a first approximation, 
this structure can be described by a straight step (see Fig.~6). 

After this consideration, a new set of factors was defined to account for 
this step-like form (Fig.~6): 
$f_1$~-- the slope between 1 and 3 Ryd; 
$f_2$~-- the depth of the \ion{He}{ii} absorption through; 
$f_3$~-- the energy of the \ion{He}{ii} ionization break; 
$f_4$~-- the depth of the \ion{He}{ii} ionization break;
$f_5$~-- the slope of the \ion{He}{ii} ionization break; 
$f_6, f_7$~-- the height and width of the
\ion{He}{ii} re-emission bump; 
$f_8$~-- the slope after the \ion{He}{ii} ionization break. 
In the following analysis, factors 
$f_3$ and $f_8$ turned out to have low effects and
were set to 4 Ryd and --0.45, respectively. 

The final UV background is shown by the short-dashed line in Fig.~6. 
The synthetic line profiles calculated with this SED are
plotted by the smooth solid lines in Fig.~5 and the corresponding 
physical parameters are listed in Table~2, Col.~1. 
The predicted column density of \ion{He}{ii} for
this absorber is $2.8\times 10^{19}$ \cm\, giving $\eta$ = 87.

In the framework of our approach, the restored spectrum may be
considered as some average spectrum. 
This spectrum
reveals features stemming from both the intergalactic incident radiation 
(depression between 3 and 4 Ryd in Fig.~6) 
and local processes in the cloud itself 
(enhanced re-emission of \ion{He}{ii} Ly$\alpha$, deep break at 4 Ryd). 
To reconstruct the true
metagalactic spectrum,  
absorbers optically thin  in \ion{He}{ii}   
are needed,
i.e. those with $N$(\ion{H}{i}) $\la 10^{15}$ \cm.

\begin{figure*}[t]
\vspace{0.4cm}
\hspace{-0.5cm}\psfig{figure=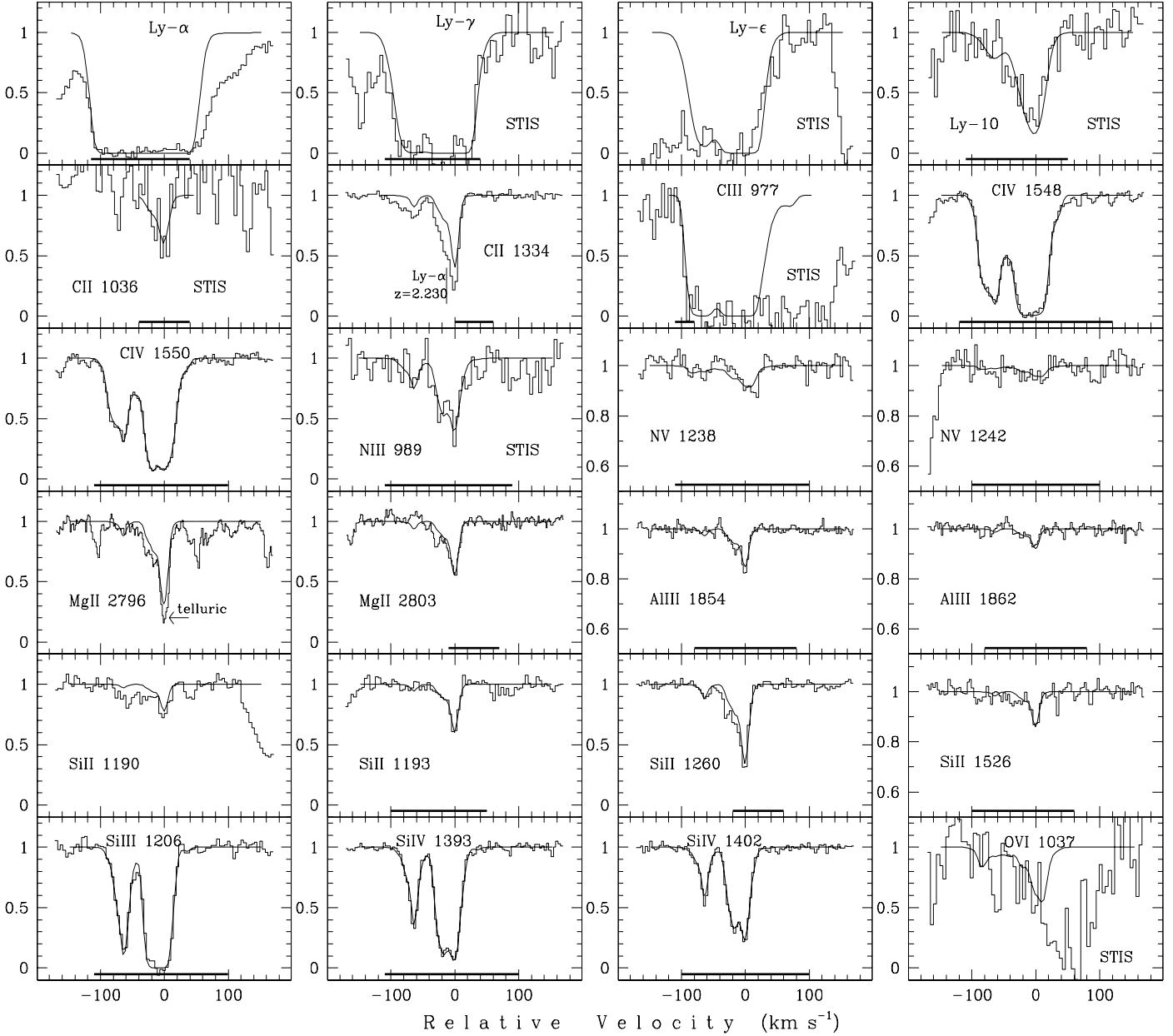,height=17.0cm,width=19.0cm}
\vspace{-0.5cm}
\caption[]{
Same as Fig.~7 but for the \zabs = 1.9426 system towards
\object{J 2233--606}.
The MCI synthetic profiles (smooth curves) are calculated with
the SEDs shown by the dotted and dashed lines in Fig.~10.
The corresponding physical parameters are listed 
in Table~2, Col.~5.
The zero radial velocity is fixed at $z = 1.9426$ 
}
\label{fig9}
\end{figure*}

\subsection {Absorption system at \zabs = 2.9659 towards \object{Q 0347--3819}}

This system has been described in detail in Levshakov et al. (2003b). 
Since then \object{Q 0347--3819} was re-observed at the VLT/UVES 
with a full wavelength coverage and longer exposure time. 
This allowed to obtain a high-quality spectrum with resolution
of $\sim$~6 \kms and S/N = $50-100$. 
Here we repeat the analysis for the new data.

The hydrogen and metal absorption lines 
of the \zabs = 2.9659 system are shown in Fig.~7 (histograms).
Neither \ion{C}{ii} $\lambda1334$ \AA\ (undetected), 
nor \ion{Si}{ii} $\lambda1260$ \AA\, 
(blended) can be used. 
This makes the presence of \ion{C}{iii} $\lambda977$ \AA\, 
line  crucial for the SED estimation since 
\ion{Si}{iii}, \ion{Si}{iv} and \ion{C}{iv} lines can be
fitted with a very broad range of ionizing spectra (constrained only by
the condition [Si/C] $< 0.5$). Although we did not find any 
metal line candidate for blending with
\ion{C}{iii}, we cannot exclude blending
with a Ly$\alpha$ forest absorption.
Thus, the analysis below should be taken with caution and it is
valid only if the absorption at the position of \ion{C}{iii} 
is entirely due to this ion.

Calculations with the HM ionizing background showed that this spectrum
significantly underestimates  \ion{C}{iii}. 
Our trials to reconstruct the
underlying continuum shape were conducted  with two 
sets of factors, with and without accounting for the
\ion{He}{ii} Ly$\alpha$ absorption. 
In both cases it was possible to obtain the appropriate ionizing spectra.
However, the spectrum estimated without \ion{He}{ii} 
Ly$\alpha$ absorption 
shows significantly enhanced emission at 3 Ryd
and at the same time 
is harder at $E > 4$ Ryd compared to the initial HM spectrum. 
Since the absorber under study has $N$(\ion{H}{i}) $= 3.4\times10^{14}$ \cm\, 
and is optically thin in \ion{He}{ii},
this type of ionizing spectrum is obviously an artifact and should be rejected. 

The recovered spectrum with the intensity depression at 
3 Ryd$ < E <$ 4 Ryd 
is shown in Fig.~6 by the point-dashed line 
with the corresponding synthetic profiles 
plotted by the smooth lines in Fig.~7 (physical parameters are given in 
Table~2, Col.~3). 
Since the depth of the \ion{He}{ii} Ly$\alpha$ trough (factor $f_2$)  
affects quite strongly  
the ion fractions, its value is estimated with accuracy better than 0.1 dex. 
The depth of the \ion{He}{ii} continuum absorption (factor $f_4$) 
has an uncertainty of about 0.15 dex.
The predicted \ion{He}{ii} column density is
$2.1\times10^{16}$ \cm\, ($\eta$ = 62) and, hence,
this cloud is optically thin in \ion{He}{ii},
$\tau^{\rm cont}_{\rm He\,II} \simeq 0.03$. 
Thus, the recovered  background coincides in this case
with the infalling radiation spectrum and is not modified by local effects.
From the depth of the \ion{He}{ii} Ly$\alpha$ trough
we can estimate the opacity of the intergalactic
\ion{He}{ii} Ly$\alpha$ absorption as 
$\tau_{{\rm He\sc\,II}}$ = $2.1\pm0.2$.

\subsection {Absorption system at \zabs = 2.9375 towards \object{HE 0940--1050}}

The absorption system at \zabs = 2.9375 towards \object{HE 0940--1050}
reveals metal lines of
\ion{C}{iii} $\lambda977$ \AA, \ion{C}{iv} $\lambda\lambda1548, 1550$ \AA\,
and \ion{O}{vi} $\lambda1037$ \AA\,
(\ion{O}{vi} $\lambda1031$ \AA\, is blended, as well as
\ion{N}{v} $\lambda\lambda1238, 1242$ \AA) (Fig.~8) .
We cannot exclude that  the observed \ion{O}{vi} $\lambda1037$ \AA\,
and \ion{C}{iii} lines are
contaminated by some hydrogen absorption. 
However, no metal line candidates for blending were found.
As in the case of the \zabs = 2.9659 system described in the preceding
subsection, the analysis below is valid only if the intensities are
entirely due to absorption of the corresponding ions.

Two lines of one element along with one line of another can be described with a 
very broad range of ionizing spectra. 
However, different spectra will deliver different abundance ratios 
[O/C] and this can be used to constrain the allowable SEDs. 
Measurements of [O/C] in Galactic
and extragalactic \ion{H}{ii} regions and in metal poor halo stars indicate that
a safe upper bound for [O/C] is 0.5 (Henry, Edmunds, \& K\"oppen 2000; 
Akerman et al. 2004).

Calculations with the HM spectrum produce for this absorption system 
[O/C] $> 1$ suggesting that
the shape of the input ionizing background could be inadequate. 
To meet [O/C] $< 0.5$,
the ionizing spectrum should be either significantly harder 
(by 0.5 dex) at $E > 4$ Ryd or should have 
a step-like depression at $E \ga $ 3 Ryd. 
The second option is preferable
since this step is also present  in the spectrum recovered from
the absorber at \zabs = 2.9171 along the same line of
sight (Sect.~3.1).  
The step at 3 Ryd $< E <$ 4 Ryd 
strongly affects the \ion{C}{iii}/\ion{C}{iv} ratio
leading to the result that the observed ratio is obtained at a higher ionization
parameter as compared to the HM spectrum and, hence, 
the corresponding \ion{O}{vi} fraction is larger.

Using the HM spectrum as an initial approximation 
and varying the depth of the \ion{He}{ii} Ly$\alpha$ absorption trough
(i.e. one-factor experiment) we adjusted 
the spectral shape in a way that enables  the description of the observed lines 
with [O/C] $< 0.5$.
The obtained spectral shape is shown in Fig.~6 (long-dashed line), 
with the synthetic profiles plotted by
the smooth lines in Fig.~ 7 and the physical parameters given in 
Table~2, Col.~4. 
The uncertainty in the step depth is of 0.1 dex 
(given the energy above 4 Ryd at the level of the HM spectrum). 
The predicted column density of the once ionized helium
$N$(\ion{He}{ii})  $ = 8.2\times10^{16}$ \cm\, 
translates to $\eta$ = 146 which is
almost 2.5 times higher as in the \zabs = 2.9659 system. 
 
\begin{figure}[t]
\vspace{0.0cm}
\hspace{0.0cm}\psfig{figure=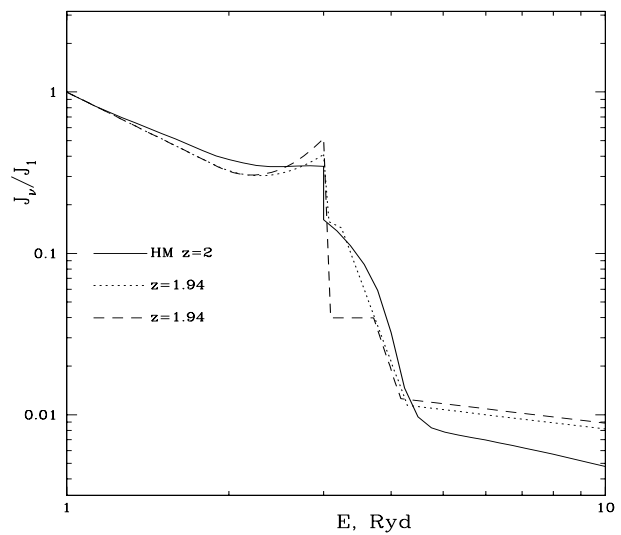,height=8.0cm,width=8.0cm}
\vspace{-0.4cm}
\caption[]{
SEDs recovered from the absorption line system at $z=1.9426$ towards
\object{J 2233--606}. For details, see text
}
\label{fig10}
\end{figure}

\begin{table*}[t]
\centering
\caption{
Physical parameters of the metal absorbers derived by the MCI procedure
(limits are given at the 1~$\sigma$ level)
}
\label{tbl-t2}
\begin{tabular}{lcccc}
\hline
\noalign{\smallskip}
Parameter$^a$ &\zabs = 2.9171&\zabs = 2.9659&\zabs = 2.9375&\zabs = 1.9426\\
 & {\scriptsize \object{HE 0940--1050}} & {\scriptsize \object{Q 0347--3819}} &
{\scriptsize \object{HE 0940--1050}} & {\scriptsize \object{J 2233--606}}\\  
(1) & (2) & (3) & (4) & (5) \\
\noalign{\smallskip}
\hline
\noalign{\smallskip}
$U_0$& 2.46E--2 & 2.26E--2 & 1.87E--1 & 1.97E--2  \\
$N_{\rm H}$, \cm& 6.00E20 & 7.22E17 & 3.19E19 & 3.20E19 \\
$\sigma_{\rm v}$, \kms & 28.4 & 14.2 & 35.2 & 33.5 \\
$\sigma_{\rm y}$& 0.75 & 0.54 & 0.70 & 0.75 \\
\noalign{\smallskip}
$Z_{\rm C}$&3.8E--7 & 1.02E--4 & 3.0E--6 & 5.3E--5 \\
$Z_{\rm N}$&$\la$1.3E--8 & 9.0E--6 & $\ldots$ & 4.9E--6 \\
$Z_{\rm O}$& $\ldots$ & $\ldots$ & 1.2E--5 & $<$\,3.0E--4 \\
$Z_{\rm Mg}$&$\ldots$ & $\ldots$ & $\ldots$ & 1.2E--5 \\
$Z_{\rm Al}$&$\la$3.8E--9 & $\ldots$ & $\ldots$ & 8.0E--7 \\
$Z_{\rm Si}$&4.9E--8 & 2.9E--5 & $\ldots$ & 9.9E--6 \\
\noalign{\smallskip}
$[Z_{\rm C}]$&--2.8 &--0.38 & --1.90 & --0.66 \\
$[Z_{\rm N}]$&$\la$--3.7 & --0.82 & $\ldots$ &--1.1 \\
$[Z_{\rm O}]$&$\ldots$ & $\ldots$ & --1.57 & $<$\,--0.2 \\
$[Z_{\rm Mg}]$&$\ldots$ & $\ldots$ & $\ldots$ & --0.44 \\
$[Z_{\rm Al}]$&$\la$--2.8 & $\ldots$ & $\ldots$ & --0.52 \\
$[Z_{\rm Si}]$&--2.8 & --0.05 & $\ldots$ & --0.51 \\
\noalign{\smallskip}
$N$(H\,{\sc i}), \cm&3.2E17 & 3.4E14 & 5.6E14& 2.5E16 \\
$N$(C\,{\sc ii}), \cm&5.8E12 & $\leq$9.0E11 & $\ldots$& 4.0E13\\
$N$(Mg\,{\sc ii}), \cm&$\ldots$ &$\ldots$ & $\ldots$ & 5.3E12 \\
$N$(Si\,{\sc ii}), \cm&8.9E11 & $\ldots$ & $\ldots$ & 5.6E12 \\
$N$(C\,{\sc iii}), \cm &2.0E14 & 4.8E13 & 5.7E12 & 1.0E15\\
$N$(N\,{\sc iii}), \cm &$\la$7.8E12 & 4.3E12 & $\ldots$ & 9.8E13\\
$N$(Al\,{\sc iii}), \cm &$\la$2.2E11&$\ldots$&$\ldots$ & 1.5E12 \\
$N$(Si\,{\sc iii}), \cm &1.2E13& 3.8E12 & $\ldots$ & 7.9E13 \\
$N$(C\,{\sc iv}), \cm &2.4E13 & 1.8E13 & 9.2E12 & 3.9E14 \\
$N$(Si\,{\sc iv}), \cm &7.6E12&6.0E12 & $\ldots$ & 5.8E13 \\
$N$(N\,{\sc v}), \cm &$\ldots$&$\ldots$ & $\ldots$ & 9.2E12 \\
$N$(O\,{\sc vi}), \cm &$\ldots$ &$\ldots$&4.2E13& $<$\,1.4E14  \\
\noalign{\smallskip}
$\langle T \rangle$, K & 2.8E4 & 1.3E4 & 3.7E4 & 1.8E4 \\
\noalign{\smallskip}
$N$(He\,{\sc ii})$^b$, \cm &2.8E19 &2.1E16 & 8.2E16 & 1.1E18 \\
$\eta^b$ & 87 & 62 & 146 & 44\\
\noalign{\smallskip}
\hline
\noalign{\smallskip}
\multicolumn{5}{l}{$^aZ_{\rm X} = N_{\rm X}/N_{\rm H}$;
$[Z_{\rm X}] = \log (N_{\rm X}/N_{\rm H}) -
\log (N_{\rm X}/N_{\rm H})_\odot$.}\\
\multicolumn{5}{l}{$^b$Predicted values.}\\
\end{tabular}
\end{table*}

\subsection {Absorption system at \zabs = 1.9426  towards \object{J 2233--606}}

This system has been described by Prochaska \& Burles (1999),
D'Odorico \& Petitjean (2001) and Levshakov et al. (2002).
A large wavelength coverage provided by combining the VLT/UVES and 
HST/STIS spectra, allows to identify many higher
order hydrogen lines and numerous metal transitions (Fig.~9). Unfortunately,
the important \ion{C}{ii} $\lambda1334$ \AA\,
line is blended with a Ly-$\alpha$
forest absorption, and \ion{C}{ii} $\lambda1036$ \AA\,
is very noisy, but this system is nevertheless worth being  analyzed due to 
the rare occasion of simultaneous presence of low- (\ion{Si}{ii}, \ion{Mg}{ii})
and high ionization lines like \ion{N}{v}. 

Calculations with the $z = 2$ 
HM ionizing spectrum (Fig.~10, solid line) describe 
most of the observed line profiles
except \ion{N}{iii} $\lambda989$ \AA\, 
and \ion{N}{v} $\lambda\lambda1238, 1242$ \AA\,
which come  over- and underestimated, respectively. 
The SED was adjusted in two
experiments with factor sets both accounting for the
Gunn-Peterson (GP) 
absorption and without it,  and
with the response maximizing the product of the ratios 
$\Upsilon_{\rm N\,V}/\Upsilon_{\rm N\,III}, 
\Upsilon_{\rm C\,II}/\Upsilon_{\rm C\,IV}$, 
and  $\Upsilon_{\rm Si\,IV}/\Upsilon_{\rm Si\,III}$.
All trials showed that to reproduce the nitrogen lines 
the ionizing background should be significantly harder
at $E > 4$ Ryd than the initial HM spectrum. Unfortunately, because of a low
quality of the \ion{C}{ii} $\lambda\lambda1334, 1036$ \AA\, lines, 
it turned out to be impossible to 
distinguish between the ionizing spectra either 
with or without the GP absorption. 
In fact,
both recovered SEDs (shown by the dotted and dashed lines in Fig.~10) give
identical fitting to the data. The synthetic profiles are shown by
the smooth lines in Fig.~9. 
However, an upper limit can be set for a
putative GP \ion{He}{ii} absorption at $z = 2$: 
$\tau_{{\rm He\,II}} < 1.8$. 

The predicted column density of \ion{He}{ii} of
$1.1\times10^{18}$ \cm\, ($\eta$ = 44) 
gives the optical depth in the \ion{He}{ii}  continuum of 1.5.
In principle, taken at face value such optical depth
can soften noticeably the
incident ionizing continuum at $E >$ 4 Ryd. 
However, the observed lines of \ion{N}{iii}
and \ion{N}{v} clearly require a hard spectrum in this energy range. 
This contradiction can be
explained in two ways: either the \ion{N}{v} lines arise in the external
regions of the absorbing cloud where the incident ionizing continuum is not
yet distorted by the \ion{He}{ii} continuum absorption, or the density
variations along the line of sight make the effective 
optical depth smaller.
In both cases 
we can conclude that the metagalactic ionizing spectrum at 
$z \sim 2$ is harder at $E > 4$ Ryd 
as compared to the spectrum predicted by HM.

\section{Conclusions}

We have proposed a method to estimate the shape of the ionizing continuum from
metal lines observed in the intervening absorption systems. The implementation 
includes the following steps:
\begin{enumerate}
\item[--] parameterization of the spectral shape by means of a set of factors 
based on physical processes relevant for the IGM;
\item[--] choice of some quantitative measure (response) to
estimate the fitness of a trial shape;
\item[--] generation of trial shapes by randomization of the factor values
according to some experimental design;
\item[--] evaluation of the factor effects;
\item[--] estimation of the optimal shape by moving in factor space towards
the ascending fitness.
\end{enumerate}
The result depends on the
number of metal lines involved in the analysis: 
in general, the more lines of different
ionic transitions of different elements are detected in an absorption
system the higher is the accuracy of the recovered spectral shape. 
In some cases additional information concerning, for instance, 
element abundance ratios can significantly tighten the 
factor values, in particular
when only a few metal lines are available. 

Although the main objective of this work is to illustrate how the
proposed approach can be used in practice, some physical results are worth
mentioning as well. 
They are as follows:
\begin{enumerate}
\item The metagalactic ionizing spectrum at redshift $z \sim 3$
has breaks at 3 and 4 Ryd and is quite hard at $E > 4$ Ryd, at least
as hard as a model metagalactic spectrum of Haardt \& Madau (1996). 
\item The intensity decrease between 3 and 4 Ryd is probably produced mostly  by 
\ion{He}{ii} Ly$\alpha$ absorption in the intergalactic diffuse gas since
line-blanketing from discrete Ly$\alpha$
forest clouds cannot account for significant \ion{He}{ii} opacity with
such hard ionizing spectrum
(Fardal, Giroux \& Shull 1998; Zheng, Davidsen \& Kriss 1998). 
Thus, the intensity depression 
at 3 Ryd $< E < 4$  Ryd in the spectrum of the
metagalactic ionizing radiation may be an imprint of a true
\ion{He}{ii} Gunn-Peterson effect. 
\item The \ion{He}{ii} opacity estimated from
the depth of the \ion{He}{ii} Ly$\alpha$ absorption trough
is $\tau_{{\rm He\,II}}$ $\sim$ 2. 
This should be considered as an average optical depth since
in our approach the \ion{He}{ii}
Ly$\alpha$ absorption is approximated by a straight step. 
As known from observations,
\ion{He}{ii} opacity at $z \sim 3$ fluctuates revealing both absorption
troughs and opacity gaps. For reference, 
$\tau_{{\rm He\,II}}$ $\sim$ 4 is measured
in absorption troughs at $2.77 < z < 2.87$  in the spectrum
of \object{HE 2347--4342} (Zheng et al. 2004) 
along with $\tau_{{\rm He\,II}} < 0.5$
in the opacity gaps at \zabs = 2.817 and 2.866 (Reimers et al. 1997).
\item A putative galactic contribution to
the ionizing background spectrum 
(e.g. Pettini et al. 2001; Ciardi, Bianchi \& Ferrara 2002; 
Fujita et al.  2003), 
if present, would significantly soften the spectrum at
$E > 4$ Ryd. We do not find any traces of soft component in 
the recovered spectra: even in the absorption systems at 
$z = 2.96$ and $z = 1.94$,
where high metallicity supposes their kinship with galaxies, the
observed line intensities point to a hard ionizing background. 
Thus, we can conclude that the ionizing background at $z \sim 3$ and $z \sim 2$
is dominated by QSOs.
\end{enumerate}

\begin{acknowledgements}
The work of I.I.A. and S.A.L. is supported by
the RFBR grant 03-02-17522 and by the RLSS grant 1115.2003.2.
\end{acknowledgements}

\appendix

\section{An example of experimental design for 7 factors}

The factors are tested at two levels:
$f_i = f_{0,i} \pm \sigma_i$, where $\sigma_i$ is a suitable scale.
Usually $\sigma_i \sim (0.1-0.2)f_{0,i}$, in order to maintain the linear
dependence of the response $\tilde{{\cal R}}$ on the factors.
The normalized and centered factor values at the upper and lower levels are,
respectively, $+1$ and $-1$. Then a 7-factorial saturated simplex design
can be
defined by the following symmetric and orthogonal matrix $\{ \hat{f}_{ij} \}$
(Nalimov 1971):

\noindent\smallskip

\begin{minipage}[t]{18.0cm}\scriptsize
\begin{tabular}{rrrrrrrr|c}
 & $\hat{f}_1$ & $\hat{f}_2$ & $\hat{f}_3$ & $\hat{f}_4$ & $\hat{f}_5$ &
$\hat{f}_6$ & $\hat{f}_7$ & $\tilde{{\cal R}}$ \\
\hline
\noalign{\smallskip}
+1&--1&--1&--1&+1&+1&+1&--1&${\cal R}_1$ \\
+1&+1&--1&--1&--1&--1&+1&+1&${\cal R}_2$ \\
+1&--1&+1&--1&--1&+1&--1&+1&${\cal R}_3$ \\
+1&+1&+1&--1&+1&--1&--1&--1&${\cal R}_4$ \\
+1&--1&--1&+1&+1&--1&--1&+1&${\cal R}_5$ \\
+1&+1&--1&+1&--1&+1&--1&--1&${\cal R}_6$ \\
+1&--1&+1&+1&--1&--1&+1&--1&${\cal R}_7$ \\
+1&+1&+1&+1&+1&+1&+1&+1&${\cal R}_8$ \\
\noalign{\smallskip}
\hline
\end{tabular}
\end{minipage}\hfill

\bigskip\noindent
The first column is added to estimate the free term $\beta$ in the model
(\ref{eq:E1b}). Such experimental design 
is called a `two level saturated plan' since here
all degrees of freedom $N = 8$ are used to estimate 8 regression coefficient
($\{ \alpha_i \}^7_{i=1}$ and $\beta$):
$$
\alpha_i = \frac{1}{N} \sum^N_{j=1} \hat{f}_{ji}{\cal R}_j\; ,
$$
and
$$
\beta = \frac{1}{N} \sum^N_{j=1} {\cal R}_j \; .
$$
Other plans with different numbers of levels and treatments (rows in the matrix)
can be used as well.

After the regression coefficients for the model (\ref{eq:E1b}) are found,
new factor values can be calculated from
$$
\hat{f}^{\rm new}_i = f_{0,i} + \Delta {\cal R}\,\sigma_i\,\mu_i\; ,
$$
where $\mu_i = \alpha_i/\sqrt{\sum^7_{j=1}\,\alpha^2_j}$
are direction cosines
and $\Delta {\cal R}$ is the increment of the response.
The value of $\Delta {\cal R}$ is restricted by the condition
$\chi^2 \sim 1$\, for every absorption line (or its portion) involved in the
optimization procedure.

\end{document}